**Presentist Fragmentalism and Quantum Mechanics**

Paul Merriam[1]

**Abstract**

This paper states and gives three applications of a novel 'presentist fragmentalist' interpretation of quantum mechanics. In a cognate paper it was explicitly shown this kind of presentism is consistent with special relativity and that it has implications for how to understand time as it relates to the Big Bang [5]. In this paper we narrowly focus on three applications. These are surely the most important conundrums for any proposed interpretation of quantum mechanics: Schrodinger's Cat, Bell non-locality, and the Born rule. It will be shown that these can be handled in a consistent and intuitive way.

**1. Introduction**

This paper gives the beginnings of a 'presentist fragmentalist' interpretation of quantum mechanics. Many aspects of the interpretation, including the philosophical arguments recommending it, must be left for cognate papers. This ambition of this paper is to merely state the interpretation and apply it to what are surely the most important conundrums for any interpretation: Schrodinger's Cat, Bell non-locality, and the Born rule. It is shown these can be handled in a consistent and intuitive way.

Two of the main ideas (to be explained below) might be stated: 1. one dimension of time in each quantum system is characterized by McTaggart's two different series: the A-series (future-present-past) and the B-series (earlier-times to later-times) [3] and 2. reality is fragmented [1] in that each quantum system does *not* contain the information of another system's A-series and thereby delineates a fragment. Reality is broken up into fragments in this way.

In a cognate paper it was explicitly shown how this interpretation is consistent with special relativity [5].

Contents

---

1    Unaffiliated, Santa Cruz, CA. pmerriam1@gmail.com

## 2. The Interpretation

This section will only state the interpretation. Many aspects must be left for cognate papers. But it will be clarified by applications in subsequent sections.

**2.1** One dimension of time is characterized by *two* series: an A-series (future-present-past) and a B-series (earlier-times to later-times) [3]. The A-series is not derivable from the B-series *in any way*. This is called an A-theory of time (see also [7]).

In this particular presentist A-theory we will assume the B-series 'goes past' (see figure 1 and section 2.6) the A-series present. As later and later times 'become' from the future into the present and then into the past, time goes on. This accords with experience.

**Figure 1**[2]

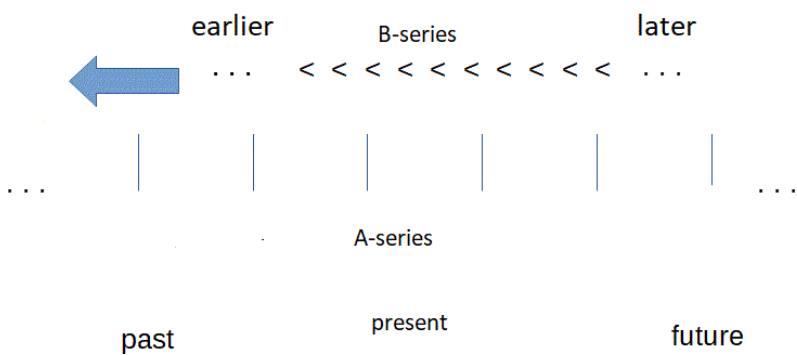

**2.2** Each quantum system has a total of *five* parameters: the A-series $\tau$, the B-series $t$, and the three spatial parameters $x^a$. This is in contrast to Minkowski space which has the *four* parameters: the B-series $t$, and the three spatial parameters $x^a$.

**2.3** Each quantum system can serve as a reference system from which one describes another quantum system in the sense of quantum reference frames. There has been somewhat of a convergence to this idea; see [2], [4], [6].

If $q_1$ is a reference quantum system from which there is another system $q_2$ in a quantum state, then $q_2$ can be taken to be a reference quantum system from which the system $q_1$ is in a quantum state. This is analogous to (but not identical with) how different frames of reference work in relativity: any frame can be taken to be the 'rest frame' from which the other frames are given by the Lorentz transformations.

**2.4** Each quantum system, no matter how small, simple, or non-local[3], forms an ontological *fragment*. In a fragment there is no fact-of-the-matter about the value of the relevant parameter in another fragment. In this interpretation the relevant parameter is the A-series.

---

2   This figure is borrowed from "A Theory of the Big Bang in McTaggart's Time," which has been accepted for publication.
3   Non-local in the spatial coordinates of another fragment.

**2.5** In each fragment there is no fact-of-the-matter about the A-series of another fragment. This is the crucial point. An A-series delineates each fragment. Reality is not a cohesive whole but is 'fragmented' in this way.

The A-series contains the information of 1. the future, 2. the present moment or 'now,' 3. the past, and 4. the notion of 'becoming' (see also [8]) for each quantum system individually. But there is no fact-of-the-matter about the A-series of two different fragments taken together. The upshot is that there is no fact-of-the-matter about a present moment in two different fragments taken together. If we have information about the present moment of one fragment $f_1$, then we do not have information about the present moment of another fragment $f_2$, and *vice versa*.

It is worth very briefly touching on the powerful argument recommending this interpretation. My subjective phenomenal experience is not accessible to you, and *vice versa*. There is no (ontological) fact-of-the-matter about whether our experiences of the color green (for example) are qualitatively the same or not. What I call 'green' is what you might call 'blue'. Meanwhile, the A-series is usually argued to also be phenomenal. So there is no (ontological) fact-of-the matter about the A-series of two distinct fragments taken together.

Further, any quantum system should be just as good as any other quantum system, so arguably one may suppose a mild form of panpsychism (only to the extent that an A-series can be defined within each fragment). The interpretation of quantum mechanics of this paper follows immediately. Of course, an adequate discussion of this philosophical argument would take this paper too far afield, so this philosophical discussion will not be pursued further here.

**2.6** Two fragments become one fragment when and only when they (mutually) interact in such a way that they become to share the same A-series, at which point they become one fragment.

This interaction is described as usual by the collapse of the state-function in a Hilbert space as defined in the parameters of the reference fragments. This collapse is taken to be described by a projector irreducibly *projecting*. That is, in some sense, it is irreducibly a *verb*. This models the large arrow in the upper left of Figure 1.

We need one more principle:

**2.7** An experimental outcome can be demonstrated/realized *only* in a fragment's present moment (as opposed to a future or a past moment). An outcome in the future, or an outcome in the past, can be discussed and theorized about. But the result of an experiment can be *demonstrated* only in the present.

**3. Schrodinger's Cat**

This section explains how the Schrodinger's Cat conundrum is handled in this interpretation.

Suppose the experimenter is Alice. The traditional paradox is that at some time during the experiment, Alice describes (so to speak) the Cat's state as a superposition, schematically as [psi> = [alive> + [dead>. Yet at that time the cat describes itself as being in one definite state, either 'alive' or else 'dead', and not in the superposition [psi>. What's going on?

The problem from the perspective of the interpretation here is that we assumed the A-series values of the Cat are the same as the A-series values of Alice during the experiment. This was sneaked in with "at *that* time". But in this theory there is no fact-of-the-matter about both the A-series (and therefore the present moment) of Alice on one hand, and the A-series (and therefore the present moment) of the Cat during the experiment on the other hand, considered together. During the experiment they are two different fragments.

But if, during the experiment, Alice and the Cat are never in a shared present moment, or a shared 'now', then *there is never a time* at which the cat gets ascribed different states, one by Alice and a different one by the Cat. That is the crucial point. And by (2.7) this means there is no fact-of-the-matter in reality about the relevant parameters of the state of the Cat in terms of the parameters of Alice, and *vice versa*, during the run of the experiment. That is how the paradox is resolved in this interpretation.

It is worth seeing in a little more detail how, if Alice can serve as a reference fragment in whose parameters the Cat can be described, then Cat can serve as a reference fragment in whose parameters Alice can be described.

Suppose that, in obvious notation, for Alice, Schrodinger's Cat is in the state

$$[psi> = c_1[alive> + c_2[dead> \qquad (1)$$

in a Hilbert space $H^{Alice}$ defined from Alice's fragment. Then, for Cat, Alice is, in obvious notation, in the corresponding state

$$[psi'>' = c_3[happy>' + c_4[sad>' \qquad (2)$$

in a Hilbert space $H^{Cat}$. Note (1) obtains in Alice's fragment if and only if (2) obtains in Cat's fragment.

The long-run statistics of the first and second terms in (1) and (2), respectively, must be the same, so in view of the Born rule we have

$$|c_3|^2 = |c_1|^2 \quad \text{and} \quad |c_4|^2 = |c_2|^2 \qquad (3)$$

The state-vector $[\Psi>$ collapses upon observation of Cat by Alice when and only when the state-vector $[\Psi'>'$ collapses upon observation of Alice by Cat.

**4. Bell Pairs and Non-locality**

The explanation of non-locality is essentially the same as with Schrodinger's Cat.

Suppose Alice and Bob are space-like separated fragments and a pair of entangled electrons, a third fragment, goes out to them. Alice decides on the orientation of her detector and then measures the spin of a relevant electron. Suppose Bob then does the same at a sufficiently long time after Alice.

For Alice, the pair does not have particularized spins (and indeed the electrons do not have particularized identities) *until* measurement. In this interpretation, that is because of the simple fact that Alice's fragment and the pair's fragment do not have the same A-series, and therefore the same present moment, or 'now', *until* such a measurement. The pair can have particular spins only in the pair's present moment. But in Alice's fragment there is no fact-of-the-matter about when the pair's present moment is. Alice's fragment does not contain that information—that is, until measurement. And *vice versa*.

The five coordinates of Alice's fragment extend throughout space. It is irrelevant how far away something is.[4] Thus, one wants to say that the non-local hidden variable of this realist interpretation is a fragment's A-series. But it should be noted that a present moment is actually not 'hidden' at all. It is one of the most fundamental empirical datum given (I am always in my present). Thus it would be more accurate to call each fragment's A-series a *non-local self-evident variable*.

## 5. Derivation of Part of the Born Rule

We'll derive the Born rule for real numbers for two systems with two possible outcomes of a mutual measurement and indicate how it is generalizable.

Let there be two quantum systems Alice and Bob (which as always could microscopic, spatially extended, etc.). In fragmental quantum mechanics these systems form ontological fragments, and a measurement (collapse of the state-function) is a mutual measurement. Alice measures Bob if and only if Bob measures Alice.

Suppose the two systems interact—they perform a measurement on each other. Let there be two possible measurement outcomes, $m_1$ and $m_2$. In Alice's fragment let $p_1$ be the 'chance' that the measurement outcome will be $m_1$, and $p_2$ be the 'chance' that the measurement outcome will be $m_2$. The measurement is explicitly not a measurement of Alice on herself. So the requirement that $p_1$ and $p_2$ sum to 1, as they would if they were normal or 'objectival' probabilities, would be *unphysical*.

Similarly, if $p_3$ and $p_4$ are the 'chances' that the measurement outcome is $m_1$ or $m_2$ respectively in Bob's fragment, it would be *unphysical* to require that they sum to 1.

Instead, we can only require that the 'chance' that Alice's gets an outcome *and* Bob's gets an outcome, sum to 1. There is a measurement outcome if and only if *both* Alice gets an outcome *and* Bob gets an outcome.

This implies that—to model the physical situation—we can only require that the *product*

$$(p_1 + p_2)(p_3 + p_4) \qquad (4)$$

sums to 1.

This gives

$$p_1 p_3 + p_1 p_4 + p_2 p_3 + p_2 p_4 = 1 \qquad (5)$$

---

[4] Note the relativity of simultaneity is a B-series feature. To incorporate the A-series of the reference system a generalization of the Lorentz transformations must be used [5].

Next, we cannot have inconsistent measurements from the two fragments. If Alice gets outcome $m_1$ then Bob must get outcome $m_1$ also, and Bob must not get outcome $m_2$. Similarly, if Alice gets outcome $m_2$ then Bob must get outcome $m_2$ also, and Bob must not get outcome $m_1$. (And *vice versa*.) This implies that for the associated 'chances' $p_1, p_2, p_3, p_4$,

$$p_1 p_4 = 0, \quad p_2 p_3 = 0 \tag{6}$$

as these would correspond to different outcomes for Alice and for Bob of the same (mutual) measurement.

Finally, Alice and Bob make use of the same theory in describing the opposing system (namely, quantum mechanics). So in the long-run statistics (i.e. tomographically) the outcome $m_1$ must have an equal chance of obtaining in both fragments, and similarly for outcome $m_2$. Thus

$$p_3 = p_1, \quad p_4 = p_2 \tag{7}$$

Applying equations (6) and (7) to equation (4) we get

$$p_1^2 + p_2^2 = 1 \tag{8}$$

for the 'chances' $p_1$ and $p_2$ in Alice's fragment, and we get the analogous equation for the 'chances' $p_3$ and $p_4$ in Bob's fragment:

$$p_3^2 + p_4^2 = 1 \tag{9}$$

This is how the Born probabilities for real numbers $p_i$ can be derived in this interpretation. The coefficients have been restricted to 'real' numbers because we wanted to interpret the $p_i$ as 'chances' in some relatively uncontroversial way for the sake of illustration.

It is clear that Alice and Bob must agree on the measurement outcome upon mutual observation. But the fact that this derivation *uses* that fact is a great virtue: for something as fundamental as the Born rule we would hope to use something fundamental in the derivation. Further, it seems that in some sense (8) and (9) only constrain the $p_i$ to be complex.

Equations (8) and (9) generalize to more than two possible outcomes of a measurement interaction. They are generalizable to *n* possible outcomes, each with a 'chance' given in each fragment, provided there are only two fragments: the reference system A and the opposing system B (or reference system B and opposing system A).

Note:

It has been verified that for 3 possible outcomes and for 4 possible outcomes, and assuming that if the 'chance' (a complex number) for an outcome is $c_i$ in Alice's fragment, and that the corresponding 'chance' for that outcome is $c_i$-bar in Cat's fragment, then the solutions to the corresponding derivation above are one-dimensional (one real parameter). It might be this can be interpreted as the phase. See associated pdf [9].

## 6. Conclusions

To summarize this realist interpretation, we have 1. for each system one dimension of time is characterized by the two McTaggertian series: the A-series and the B-series. 2. each quantum system forms an ontological fragment: one system does not have the information about the A-series of an opposing system. 3. a (mutual) measurement happens when and only when the two systems come to have the same A-series.

A challenge for any proposed interpretation of quantum mechanics is to account for what are surely the three most fundamental conundrums: Schrodinger's Cat, Bell non-locality, and the Born rule. The novel interpretation of this paper accounts for these conundrums in a consistent and intuitive way. This interpretation meets the challenge successfully.